\documentstyle[aps,pre,twocolumn,floats]{revtex}
\begin{document}

\newcommand{\gsim}{
\,\raisebox{0.35ex}{$>$}
\hspace{-1.7ex}\raisebox{-0.65ex}{$\sim$}\,
}

\newcommand{\lsim}{
\,\raisebox{0.35ex}{$<$}
\hspace{-1.7ex}\raisebox{-0.65ex}{$\sim$}\,
}

\newcommand{\fractional}{ \mbox{frac} } 
\newcommand{\integer}{ \mbox{int} }

\bibliographystyle{prsty}

%
%
%

\title{
\begin{flushleft}
{\small 
PHYSICAL REVIEW E 
\hfill
VOLUME  {\normalsize 55}, NUMBER  {\normalsize 3}
\hfill 
MARCH {\normalsize 1997, 2569-2572} 
}\\
\end{flushleft}  
Quantum thermoactivation of nanoscale magnets
}

\author{D.~A. Garanin
\renewcommand{\thefootnote}{\fnsymbol{footnote}}
\footnotemark[1]
}

\address{
I. Institut f\"ur Theoretische Physik, Universit\"at Hamburg,
Jungiusstr. 9, D-20355 Hamburg, Germany \\
\smallskip
{\rm(Received 22 May 1996)}
\bigskip\\
\parbox{14.2cm}
{\rm
The integral relaxation time describing the thermoactivated escape of a 
uniaxial quantum spin system interacting with a boson bath  
is calculated analytically in the whole temperature 
range.
For temperatures $T$ much less than the barrier height $\Delta U$, 
the level quantization near the top of the barrier and the strong 
frequency dependence of the one-boson transition probability
can lead to the regularly spaced deep minima of the thermoactivation 
rate as a function of the magnetic field applied along the $z$ axis.
[S1063-651X(97)13203-2]
\smallskip
\begin{flushleft}
PACS number(s): 05.40.+j, 75.50.Tt
\end{flushleft}
} 
} 
\maketitle
\renewcommand{\thefootnote}{\fnsymbol{footnote}}
\footnotetext[1]{ Electronic address: garanin@physnet.uni-hamburg.de }

\vspace{-5mm}

The problem of the escape rate of a uniaxial magnetic particle has remained 
in the focus of attention since the work of N\'{e}el \cite{nee49}, who 
stressed the role of thermal agitation.
During the last years the interest in this problem has increased in view 
of possible applications to the information storage and in connection 
with the magnetic quantum tunneling 
(MQT; see e.g., Ref. \cite{stachubar92}).

For classically large particles, the thermoactivation escape rate  
was first calculated by Brown \cite{bro63}, 
who derived the Fokker-Planck equation 
for an assembly of particles and solved it
perturbatively in the high-temperature case, $\Delta U \ll T$, 
and with the use of the Kramers transition-state method \cite{kra40} 
for $ T \ll \Delta U$.  
In these both limiting cases the time dependence of the 
average magnetization is a single exponential, 
and the relaxation rate of ferromagnetic particles is given by the 
lowest eigenvalue $\Lambda_1$ of the Fokker-Planck equation.
For $T\sim \Delta U$ the latter is no longer the case, and the best 
measure of the relaxation rate is the 
integral relaxation time $\tau_{\rm int}$ determined as the area 
under the magnetization relaxation curve 
after a sudden infinitesimal change of the longitudinal magnetic field
\cite{garishpan90,cofetal94,cofcrokalwal95,gar96pre}.
The quantity $\tau_{\rm int}$  
can be calculated analytically in the whole range of temperatures, 
and $\tau_{\rm int}^{-1}$ coincides with $\Lambda_1$ 
in the asymptotic regions.

With the miniaturization of the magnetic particle 
both the thermoactivation and the MQT escape rates increase; the latter 
becomes dominant below the crossover temperature $T_0$
determined by the 
interactions noncommuting with the operator $S_z$ and hence causing the 
MQT.
For information storage applications the most important are 
systems with small tunneling interactions and   
correspondingly low $T_0$. 
In this case the tunneling level splittings can be calculated 
perturbatively \cite{gar91jpa} for the arbitrary spin values $S$, which 
can be advantageous in comparison to the semiclassical instanton method 
(see, e.g., Ref. \cite{stachubar92}) for nanoscale systems 
with {\em moderately} high spin values as the recently 
synthesized Mn clusters having
$S=10$ and 12 in the ground state \cite{sesgatcannov93,gatcanparses94}.
For such systems with not too large $S$ and low $T_0$ 
the thermoactivation escape rate can be 
dominant down to the temperatures where it changes due to the spin level 
quantization.

The purpose of this paper is to calculate the thermoactivation escape rate 
of a quantum spin system in terms of the integral relaxation time
starting from a spin-bath Hamiltonian of the type
\begin{eqnarray}\label{sphham}
&&
{\cal H} = - HS_z - DS_z^2 + {\cal H}_b \\
&&\qquad
{}-
\sum_{\bf q} V_{\bf q}(\bbox{\eta}{\bf S}) (a_{\bf q}^\dagger + a_{-\bf q})
-
\sum_{\bf pq} \mbox{\it V}_{\bf pq}(\bbox{\eta}{\bf S}) 
a_{\bf p}^\dagger a_{\bf q} ,
\nonumber 
\end{eqnarray}
where the arbitrary vector $\bbox{\eta}$ is for simplicity set to be 
$\eta_x=\eta_y=\eta_z=1$.
If the bath excitations described by the operators $a^\dagger$ and
$a$ are phonons, 
then the coupling to the bath that is linear in spin variables 
is prohibited by the time-reversal symmetry.
This means that
modulations of the crystal field by phonons do not produce a 
{\em fieldlike} perturbation on a spin system, and hence $\bbox{\eta}=0$.
Thus, in this case it would be better to write quadratic terms of the type 
$S_\alpha S_\beta$ instead of $(\bbox{\eta}\bf S)$ in (\ref{sphham})
(see, e.g., Ref. \cite{garkim91}).
Nevertheless, we will use (\ref{sphham})  
with ${\cal H}_b$ and the couplings $V$ 
characteristic for phonons since it is 
the most suitable for the first 
presentation of the method and describes the main qualitative features 
of the relaxation of a spin system.
Moreover, Hamiltonian (\ref{sphham}) means the direct quantum 
generalization of the Langevin-{\em field} formalism used by Brown 
\cite{bro63}
(which is subject to the same criticism), and thus provides a link 
to the known results for classical ferromagnetic particles.

If the spin-bath coupling is weak, the equation of motion for the 
density matrix of the spin system can be obtained in the second order of 
perturbation theory, and the diagonal part of it is
the well-known system of the kinetic balance equations
for the occupation numbers $N_m$ of the spin states $|m\rangle$
\begin{eqnarray}\label{keq}
&&
\dot N_m = l_m^2 ( W_{m,m+1} N_{m+1} - W_{m+1,m} N_m ) \nonumber \\
&&\qquad
{} + l_{m-1}^2 ( W_{m,m-1} N_{m-1} - W_{m-1,m} N_m ) .
\end{eqnarray}
Here $l_m=\sqrt{S(S+1)-m(m+1)}$ are the matrix elements of the operators 
$S_\pm$, and the "spin-free" transition probabilities $W$ are given by
\begin{equation}\label{wsum}
W_{m+1,m} = W(\omega_{m+1,m}) = W^{(1)} + W^{(2)},
\end{equation}
where
\begin{equation}\label{om}
\omega_{m+1,m} \equiv \varepsilon_{m+1} - \varepsilon_m = - H - D(2m+1) 
\end{equation}
are the transition frequencies and
$\varepsilon_m = -Hm -Dm^2$ are the spin energy levels,
\begin{eqnarray}\label{w1}
&&
W^{(1)}(\omega) = \sum_{\bf q} |V_{\bf q}|^2 
\big\{ (n_q+1) \pi\delta(\omega_q+\omega) \nonumber \\
&&\qquad\qquad\qquad\qquad
{} + n_q \pi\delta(\omega_q-\omega) \big\}
\end{eqnarray}
is the contribution of the one-phonon emission and absorption processes,
and
\begin{equation}\label{w2}
W^{(2)}(\omega) = \sum_{\bf pq} |V_{\bf pq}|^2
n_p(n_q+1)\pi\delta(\omega_p-\omega_q-\omega)
\end{equation}
is that of the two-phonon (Raman scattering) ones.
One can check that the transition probabilities satisfy the detailed 
balance condition: $W(\omega)=W(-\omega)\exp(-\omega/T)$, which 
ensures the static distribution of the form
\begin{equation}\label{n0}
N_m^{(0)} = \frac{1}{Z}\exp(-\varepsilon_m/T),
\qquad Z =\!\! \sum_{m=-S}^S \!\!\exp(-\varepsilon_m/T) .
\end{equation}
The quantities $W(\omega)$ with $\omega>0$ describing transitions to 
upper energy levels become exponentially small at low temperatures.
For spin-phonon couplings of the type
\begin{equation}\label{vcoupl}
V_{\bf q} \sim \theta_1 \left(\frac{\omega_q}{\Omega}\right)^{1/2},
\qquad
V_{\bf pq} \sim \theta_2 \frac{ (\omega_p\omega_q)^{1/2} }{\Omega},
\end{equation}
where $\Omega=Mc^2$, $M$ is the unit cell mass,
$\omega_q=cq$, and $c$ is the phonon 
velocity, the estimation of $W(\omega)$ with $\omega<0$ 
with the help of (\ref{w1}) and (\ref{w2}) yields (cf. Ref. \cite{orb61})
\begin{equation}\label{w1res}
\renewcommand{\arraystretch}{1.2}
W^{(1)} \sim \frac{\theta_1^2|\omega|^3}{\Theta^4}(n_{|\omega|}+1)
\cong 
\left\{
\begin{array}{ll}
\theta_1^2\omega^2 T/\Theta^4, & |\omega|\ll T
\\
\theta_1^2|\omega|^3/\Theta^4, & T\ll |\omega|
\end{array}
\right.
\end{equation}
and 
\begin{equation}\label{w2res}
\renewcommand{\arraystretch}{1.2}
W^{(2)} \sim 
\left\{
\begin{array}{ll}
\theta_2^2 \theta_D^5 T^2/\Theta^8, & |\omega|\ll \theta_D \ll T 
\\
\theta_2^2 T^7/\Theta^8, & |\omega|\ll T \ll \theta_D
\\
\theta_2^2 T^4 |\omega|^3/\Theta^8, & T \ll |\omega| \ll 
                                                              \theta_D.
\end{array}
\right.
\end{equation}
Here $\theta_D\sim\hbar\omega_{q_{\rm max}}$ is the Debye temperature,   
$\Theta^4\equiv\Omega\theta_D^3 \sim \hbar^3\rho^2 c^5$, and $\rho$ is the 
density of the lattice.

It is convenient to introduce the reduced variables
\begin{equation}\label{xial}
\xi \equiv \frac{SH}{T}, 
\qquad \alpha \equiv \frac{S^2D}{T},
\qquad h \equiv \frac{\xi}{2\alpha} = \frac{H}{2SD},
\end{equation}
which are equivalent to those used for the description of classical 
single-domain magnetic particles \cite{bro63,garishpan90} and should be 
kept constant if $S\to\infty$.
In this limit the transition frequencies $\omega_{m+1,m}$ of 
Eq. (\ref{om}) tend to zero and, accordingly, the frequency-independent 
two-phonon transition probabilities $W^{(2)}$ given by (\ref{w2res}) 
govern the relaxation.
Since the occupation numbers of the neighboring levels in (\ref{keq}) 
become close to each other, Eq. (\ref{keq}) 
goes over to the classical Fokker-Planck equation 
\cite{bro63,garishpan90,cofetal94,cofcrokalwal95}.
An extreme quantum case is realized for a three-level 
system with a barrier ($S=1$), for which the relaxation between the two 
lowest levels through the highest one (the so-called ``resonance 
fluorescence'') has at low temperatures an exponentially small rate 
$\Gamma \sim \exp(-\Delta/T)$ \cite{orb61}.

The relaxation of any initial state described by the system of the 
first-order differential equations (\ref{keq}) is described in general, 
and particularly at intermediate temperatures, 
by $2S$ exponentials of the type 
$A_i \exp(-\Lambda_i t)$, where $\Lambda_i$ are the
corresponding eigenvalues.
In such situations the best measure of the relaxation rate is the 
integral relaxation time $\tau_{\rm int}$ determined as the area under the 
magnetization relaxation curve 
after a sudden infinitesimal change of the applied field $H$: 
\begin{equation}\label{taudef}
\tau_{\rm int} \equiv \int_0^\infty \!dt 
\frac
{ m_z(\infty) - m_z(t) }
{ m_z(\infty) - m_z(0) }.
\end{equation}
Here $m_z\equiv \langle S_z \rangle/S$ is given at equilibrium by 
$m_z=B(\xi,\alpha)=\partial\ln Z /\partial\xi$.
One can calculate $\tau_{\rm int}$ analytically 
at 
arbitrary temperatures from equations (\ref{keq}) producing the 
low-frequency expansion of the linear longitudinal dynamic susceptibility 
\cite{garishpan90}
\begin{equation}\label{chitau}
\chi_z(\omega) \cong \chi_z (1 + i\omega\tau_{\rm int} + \ldots) .
\end{equation}
Here $\chi_z\equiv \partial m_z/\partial(SH)=B'/T$
is the static susceptibility, and
$B'\equiv\partial B(\xi,\alpha)/\partial\xi$.
Taking into account that the alternating field 
$\Delta H_z(t)=\Delta H_{z0}\exp(-i\omega t)$ modulates the transition 
frequencies (\ref{om}), using the detailed balance condition and 
introducing $N_m \cong N_m^{(0)}(1+Q_m)$ with 
$Q_m=q_m(\omega) S\Delta H_z(t)/T$ one comes in the linear approximation in 
$\Delta H_z(t)$ to the system of equations
\begin{eqnarray}\label{qeq}
&&
l_m^2 W_{m+1,m} ( q_{m+1} - q_m ) + 
l_{m-1}^2 W_{m-1,m} ( q_{m-1} - q_m ) \nonumber \\
&&\qquad
{}+ i\omega q_m = (l_m^2 W_{m+1,m}-l_{m-1}^2 W_{m-1,m})/S .
\end{eqnarray}
The susceptibility $\chi_z(\omega)$ can now be written as
\begin{equation}\label{chiom}
\chi_z(\omega) = \frac{1}{TS}\sum_{m=-S}^S m N_m^{(0)} q_m(\omega).
\end{equation}
The second-order finite-difference equation (\ref{qeq}) should be solved
perturbatively in $\omega$; this can be done analytically since the 
first line of (\ref{qeq}) contains only the differences of $q_m$, 
and the order of (\ref{qeq}) can thus be lowered to 1.
In the static limit, (\ref{qeq}) reduces to 
$
q_{m+1}^{(0)}-q_m^{(0)}=1/S
$
with the solution 
$
q_m^{(0)} = m/S - B
$.
In first order in $\omega$ one can introduce 
$
p_m\equiv l_m^2 W_{m+1,m}\exp(-\varepsilon_m/T) 
[q_{m+1}^{(1)}-q_m^{(1)}]
$, 
satisfying 
$
p_m-p_{m-1}=-i\omega q_m^{(0)}\exp(-\varepsilon_m/T)
$.
Finding $p_m$ and then $q_m^{(1)}$, and using (\ref{chiom}) and 
(\ref{chitau}), one obtains the final result
\begin{equation}\label{tau}
\tau_{\rm int} = \frac{1}{B'}\sum_{m=-S}^{S-1}
\frac{ \Phi_m^2 }{ N_m^{(0)} l_m^2 W_{m+1,m} } ,
\end{equation}
where $N_m^{(0)}$ is given by (\ref{n0}) and
\begin{equation}\label{phi}
\Phi_m = \sum_{k=-S}^m \left( B - \frac{k}{S} \right) N_k^{(0)}.
\end{equation}

The formulas above are valid at all temperatures, and they are 
a direct quantum 
generalization of the classical results of Ref. \cite{garishpan90}, which 
are recovered in the limit $S\to\infty$.
At high temperatures where $\alpha,\xi\ll 1$ and $W^{(1)}\ll W^{(2)}$, the 
calculation in (\ref{phi}) and (\ref{tau}) yields 
\begin{equation}\label{tauht}
\tau_{\rm int}^{-1} \cong \Lambda_N 
\left[ 
1 - \frac{2\alpha}{5}
\left(1-\frac{1}{2S}\right)\left(1+\frac{3}{2S}\right)
\right]
\end{equation}
with $\Lambda_N\equiv 2W^{(2)}$ (cf. Ref. \cite{bro63}).
At $\alpha,\xi\sim 1$ (i.e., $T\sim\Delta U$)
the sums in (\ref{phi}) and (\ref{tau}) should be performed numerically.
The recent numerical calculations of Ref. \cite{cofetal94} for the 
classical model 
have shown that, in the unbiased case, $\xi=H=0$, the difference between 
$\tau_{\rm int}^{-1}$ and the lowest eigenvalue of the 
Fokker-Planck equation 
$\Lambda_1$ does not exceed 1.2\% in the whole range of temperatures.
$\Lambda_1$ remained in the focus of interest since the work of Brown 
\cite{bro63}, but for $T\sim\Delta U$ it has no direct physical 
meaning and cannot be represented by a closed analytical formula.
One can expect that the quantities $\tau_{\rm int}^{-1}$ and $\Lambda_1$ are 
even 
closer to each other in the unbiased {\em quantum} case, since the 
difference between them disappears for $S=1/2$.

At low temperatures $\alpha\gg 1$, the thermoactivation escape rate of 
the particle becomes exponentially small.
Here, as was recently discovered in Ref. \cite{cofcrokalwal95},
$\tau_{\rm int}^{-1}\gg \Lambda_1$ for sufficiently strong bias 
($h \gsim 0.2$).
This effect was physically explained \cite{gar96pre} 
as resulting from the depletion of 
the upper potential well and the competition 
in the integral relaxation time $\tau_{\rm int}$
of the overbarrier thermoactivation with the rate $\Lambda_1$ 
and the fast relaxation inside the lower well.
These two relaxation mechanisms can be analytically separated for 
$S\gg 1$ since the summand of (\ref{tau}) consists for $\alpha,\xi\gg 1$ 
of two peaks centered at the barrier top and in the lower well 
($m\sim S$).
The barrier contribution $\tau_{{\rm int},B}$ can be related to 
$\Lambda_1$ taking into account the depletion effect \cite{gar96pre};
in the small bias case, $h\ll 1$, which will be 
considered henceforth, one obtains
\begin{equation}\label{lam1tau}
\Lambda_1^{-1} \cong \tau_{{\rm int},B} B'(\xi,\alpha) \cosh^2\xi .
\end{equation}
For $\xi=0$ one has $B'\cong 1$ and $\Lambda_1\cong \tau_{\rm int}^{-1}$.
Since $N_k^{(0)}$ in (\ref{phi}) is strongly peaked 
at low temperatures in the wells 
($k\sim \pm S$), and is small elsewhere, the function $\Phi_m$ is 
independent on $m$ and given by $\Phi_m\cong 1/(2\cosh^2\xi)$ in the 
main part of the phase space, including the barrier region.
Then (\ref{lam1tau}) and (\ref{tau}) can be combined to 
\begin{equation}\label{lam1}
\Lambda_1 \cong \frac{4S(S+1)\cosh^2\xi}{Z(\xi,\alpha)}
\left[
\sum_{m=-\infty}^\infty \frac{ \exp(\varepsilon_m/T) }{ W_{m+1,m} }
\right]^{-1} ,
\end{equation}
where the exponential factor 
$\sim 1/N_k^{(0)}$ in (\ref{lam1}) cuts the sum actually at 
$m\sim S\alpha^{-1/2}\ll S$ [see (\ref{om}) and (\ref{xial})].
The partition function $Z$ in (\ref{lam1}) for $h\ll 1$ is 
given by
\begin{equation}\label{zfunc}
\renewcommand{\arraystretch}{1.2}
Z \cong 
\left\{
\begin{array}{ll}
(S/\alpha)e^\alpha\cosh\xi , & SD \ll T \ll S^2D 
\\
2 e^\alpha\cosh\xi , & T \ll SD ,
\end{array}
\right.
\end{equation}
where $SD$ is the level spacing in the wells.

The sum in (\ref{lam1}) depends on the relation between temperature $T$
and the level spacing near the top of the barrier $\sim D$, as well as 
on the role played by the one- and two-phonon probabilities $W^{(1)}$ 
and $W^{(2)}$.
For not too low temperatures this sum can be approximated by the 
integral over $x\equiv m/S$, 
and the quantity $W^{(1)}$ of (\ref{w1res}) can be in this case 
represented in the form
\begin{equation}\label{w1bar}
W^{(1)}_{m+1,m} \cong \bar W^{(1)} \frac{\omega_{m+1,m}^2}{D^2},
\qquad \bar W^{(1)} \sim \frac{\theta_1^2 D^2 T}{\Theta^4} ,
\end{equation}
whereas $W^{(2)}\propto T^7$ is given for $T\ll\theta_D$ 
by the second line of (\ref{w2res}).
One can see that the integral over $x$ in (\ref{lam1}) 
can be cut either by the exponential function at 
$\Delta x\sim \alpha^{-1/2}$ or by the denominator 
$W(\omega_{m+1,m})$ at $\Delta x\sim S^{-1} \sqrt{ W^{(2)}/\bar W^{(1)} }$.
Crossover between these two regimes occurs at the temperature 
\begin{equation}\label{t12}
T_{12} \sim \Theta \left(\frac{D}{\Theta}\right)^{1/5} 
\left(\frac{\theta_1}{\theta_2}\right)^{2/5} 
\qquad (T_{12}\ll\theta_D) .
\end{equation}
Taking (\ref{zfunc}) into account, one obtains  
\begin{equation}\label{lam1cont}
\Lambda_1 \cong A 
\big\{ \exp[-\alpha(1+h)^2] + \exp[-\alpha(1-h)^2] \big\},
\end{equation}
where $h\ll 1$, and the prefactor $A$ is given by
\begin{equation}\label{apref}
\renewcommand{\arraystretch}{1.2}
A \cong 
\left\{
\begin{array}{ll}
2W^{(2)}\pi^{-1/2}\alpha^{3/2}, 
& SD, T_{12} \ll T 
\\
2S\sqrt{ \bar W^{(1)} W^{(2)} } \pi^{-1}\alpha , 
& SD \ll T \ll T_{12}
\\
SW^{(2)}\pi^{-1/2}\alpha^{1/2},
& T_{12} \ll T \ll SD
\\
S^2\sqrt{ \bar W^{(1)} W^{(2)} } \pi^{-1} ,
& T \ll T_{12}, SD .
\end{array}
\right.
\end{equation}
The first of these expressions coincides with that of Brown \cite{bro63}
in the high-barrier limit, if one introduces $\Lambda_N\equiv 2W^{(2)}$.
The temperature dependences of $A$ read
$T^{11/2}$, $T^{3}$, $T^{13/2}$, and $T^{4}$, respectively.

The continuous approximation above is only valid if 
$\Delta x \sim S^{-1}\sqrt{ W^{(2)}/\bar W^{(1)} }\gg S^{-1}$, 
or $\Delta m \gg 1$.
This condition sets one more charakteristic temperature
\begin{equation}\label{tq}
T_q \sim \Theta \left(\frac{D}{\Theta}\right)^{1/3}
\left(\frac{\theta_1}{\theta_2}\right)^{1/3} \qquad (T_q\ll\theta_D),
\end{equation}
below which the level quantization becomes essential.
In the temperature interval $D\ll T \lsim T_q$ the sum in (\ref{lam1})
converges at $\Delta m\sim 1$, whereas the exponential factors 
can still be neglected.
In this case $\Lambda_1$ shows a sinusoidal dependence on the 
longitudinal magnetic field in the {\em weak-field} region $H\sim D$, 
the amplitude of which is exponentially small for 
$W^{(2)}\gsim\bar W^{(1)}$ but becomes great for $W^{(2)}\ll\bar W^{(1)}$.
In the range $D \ll T \lsim T_q, SD$ one obtains formula (\ref{lam1cont}),
with the prefactor given by
\begin{equation}\label{aprefsin}
A \cong S^2 
\left\{
\frac{4}{\pi^2}\sin^2 \!\!
\left[ \pi \left( \frac{H}{2D}\! +\! S\! +\! \frac{1}{2} \right) \right] 
\bar W^{(1)} + W^{(2)}
\right\} ,
\end{equation}
This result shows deep 
minima of the thermoactivation escape rate $\Lambda_1$ for 
$H/(2D) = \pm 1/2, \pm 3/2, \ldots$ for $S$ integer and for
$H/(2D) = 0, \pm 1, \pm 2 \ldots$ for $S$ half-integer.
At such fields two levels near the barrier top become degenerate, and 
the leading contribution to the transition probability between them, 
$W^{(1)}$, disappears.

In the extreme quantum temperature region $T \ll D$, the 
sum in (\ref{lam1}) is again  determined by the exponential factor. 
Here, however, only one or maximally two terms of the sum contribute to 
(\ref{lam1}).
If there is an energy level with $m=m_{\rm max}$
just at the top of the barrier, 
like for $S$ integer and $H=0$ ($m_{\rm max}=0$), 
then equal contributions to (\ref{lam1}) 
come from the terms with $m=m_{\rm max}$ and $m=m_{\rm max}-1$,
which describe transitions between three upper levels with 
$m=m_{\rm max}, m_{\rm max}\pm 1$. 
If there are two degenerate levels to the right and left from the 
barrier top, like for $S$ half-integer and $H=0$, then the leading 
contribution is due to the term in (\ref{lam1}) describing transitions 
between these two levels.
The general result is given by formula (\ref{lam1cont}) with
\begin{eqnarray}\label{aprefq}
&&
A \cong S(S+1)  
\exp(-\Delta\varepsilon_{m_{\rm max}}/T ) \nonumber \\
&&\qquad
\times\left[
\frac{1}{ W_{m_{\rm max}+1, m_{\rm max}} } 
+
\frac{1}{ W_{m_{\rm max}-1, m_{\rm max}} }
\right]^{-1} , 
\end{eqnarray}
where $m_{\rm max} = - H/(2D) + F$,
$F\equiv{\cal F}[H/(2D) + S + 1/2] - 1/2$,
$\Delta\varepsilon_{m_{\rm max}} = DF^2$  
is the mismatch between 
$\varepsilon_{m_{\rm max}}$ and the barrier top, and 
${\cal F}(X)$ is the fractional part of $X$.
The transition probabilities $W$ in (\ref{aprefq}) are dominated by 
$W^{(1)}\propto |\omega|^3$ [see (\ref{w1res})] with 
$\omega_{m_{\rm max}\pm1, m_{\rm max}} = -D (1 \pm 2F)$,
and the escape rate $\Lambda_1$ shows qualitatively the same magnetic 
field dependence as that given by (\ref{aprefsin}).
Note that in this low-temperature limit the prefactor $A$ is 
temperature-independent, excluding the narrow field regions where it is 
determined by $W^{(2)}$ and is very small.

Summarizing, the thermoactivation escape rate of a quantum ferromagnetic 
particle was calculated microscopically in the whole temperature range, 
allowing for the frequency dependence of the transition probabilities 
and for the quantization of the energy levels.
Even in the low-temperature range, where the escape rate is exponentially 
small, the situation is determined by several characteristic energies 
and there are rather many limiting cases for the prefactor $A$ in 
(\ref{lam1cont}), which can be difficult to observe if the 
corresponding temperature intervals are not wide enough.
Not trying to give numerical estimations, since the Hamiltonian 
(\ref{sphham}) is only a schematic one suitable mainly for the 
presentation of the method and for a qualitative analysis, we make only 
some general remarks about the results obtained.
First, the greatest variety of different temperature intervals can be 
realized for particles containing a macroscopically large number $N$ of 
magnetic ions and having an effective spin $S_{\rm eff} \sim N\gg 1$. 
However, only the classical case [the first line of (\ref{apref})] 
can be practically observed, 
whereas in other cases $\Lambda_1$ is unmeasurably small due to the too 
large values of $\alpha$.

Better candidates for searching for the nonclassical 
thermoactivation rates predicted here are nanoscale systems such as 
Mn$_{12}$ clusters with $S=10$ having a strong uniaxial 
anisotropy ($\Delta U = S^2 D = 61$~K) \cite{sesgatcannov93}.
These clusters show a superparamagnetic behavior, and  
for $2~{\rm K} \leq T \leq 8~{\rm K}$ the prefactor $A$
in (\ref{lam1cont}) is temperature independent.
The latter corresponds to the extreme quantum case (\ref{aprefq}), 
which could, however, be expected only for $T\lsim D \sim 0.6~$K.
In the main range of temperatures $T\lsim S^2 D = 61$~K, one 
cannot expect one of the pure limiting forms of $A$, and should 
resort to using (\ref{lam1}) because $S=10$ is not large enough.
Nevertheless, one qualitative feature always remains: If the relaxation 
is governed by the two-phonon processes, one can expect a strong 
temperature dependence of $A$, and, if the one-phonon processes are 
dominant, then the temperature dependence of $A$ is weak or absent, but 
there should be a strong dependence on the magnetic field of the type 
(\ref{aprefsin}). 

\smallskip

The author thanks Hartwig Schmidt for valuable comments.
The financial support of Deutsche Forschungsgemeinschaft 
under contract Schm 398/5-1 is greatfully acknowledged.

\vspace{-0.5cm}


\begin{thebibliography}{10}
\vspace{-1.5cm}

\bibitem{nee49}
{L. N\'{e}el}, Ann. Geophys. {\bf 5},  99  (1949).

\bibitem{stachubar92}
{P. C. E. Stamp, E. M. Chudnovsky, and B. Barbara}, Int. J. Mod. Phys. B {\bf
  6},  1355  (1992).

\bibitem{bro63}
{W. F. Brown, Jr.}, Phys. Rev. {\bf 130},  1677  (1963).

\bibitem{kra40}
{H. A. Kramers}, Physica {\bf 7},  284  (1940).

\bibitem{garishpan90}
{D. A. Garanin, V. V. Ishchenko, and L. V. Panina}, Teor. Mat. Fiz. 
  {\bf 82}, 242 (1990)
   [Theor. Math. Phys. (USSR) {\bf 82},  169  (1990)].

\bibitem{cofetal94}
{W. T. Coffey, D. S. F. Crothers, Yu. P. Kalmykov, E. S. Massawe, and J. T.
  Waldron}, Phys. Rev. E {\bf 49},  1869  (1994).

\bibitem{cofcrokalwal95}
{W. T. Coffey, D. S. F. Crothers, Yu. P. Kalmykov, and J. T. Waldron}, Phys.
  Rev. B {\bf 51},  15947  (1995).

\bibitem{gar96pre}
{D. A. Garanin}, Phys. Rev. E  {\bf 54}, 3250 (1996).

\bibitem{gar91jpa}
{D. A. Garanin}, J. Phys. A {\bf 24},  L61  (1991).

\bibitem{sesgatcannov93}
{R. Sessoli, D. Gatteschi, A. Caneschi, and M. A. Novak}, Nature {\bf 365},
  141  (1993).

\bibitem{gatcanparses94}
{D. Gatteschi, A. Caneschi, L. Pardi, and R. Sessoli}, Science {\bf 265},  
1054 (1994).

\bibitem{garkim91}
{A. Garg and G.-H. Kim}, 
Phys. Rev. Lett. {\bf 63}, 2512 (1989);
Phys. Rev. B {\bf 43},  712  (1991).

\bibitem{orb61}
{R. Orbach}, Proc. R. Soc. London Ser. A {\bf 264},  458  (1961).

\end{thebibliography}

\end{document}